\documentclass[
  reprint,
  amsmath,amssymb,
  aps,
  prl,
  superscriptaddress,
  nobibnotes
]{revtex4-2}

\usepackage{graphicx}   % figures
\usepackage{dcolumn}    % decimal-aligned table columns
\usepackage{bm}         % bold math symbols
\usepackage{amsmath}
\usepackage{tikz}
\usepackage{xcolor}
\usepackage{comment}
\usepackage{hyperref}   % keep last among these
\hypersetup{colorlinks=true, allcolors=blue}

\def\eq#1{(\ref{eq:#1})}

\def\d{\partial}
\def\eps{\epsilon}         

\def\UHP{\mathrm{UHP}}
\def\L{\mathcal{L}}

\newcommand{\pdfsecit}[1]{\pdfbookmark[1]{#1}{#1}\textit{#1}}

\DeclareMathAlphabet      {\mathit}{OT1}{cmr}{m}{it}

\begin{document}

\title{D-brane tension as central charge}

\author{Vin\'{\i}cius Bernardes}
\email{viniciusbernsilva@gmail.com}
\affiliation{CEICO, FZU - Institute of Physics of the Czech Academy of Sciences, 
	No Slovance 2, 182 21, Prague 8, Czech Republic}

\author{Theodore Erler}
\email{tchovi@gmail.com}
\affiliation{CEICO, FZU - Institute of Physics of the Czech Academy of Sciences, 
	No Slovance 2, 182 21, Prague 8, Czech Republic}

\author{Atakan Hilmi F{\i}rat}
\email{ahfirat@ucdavis.edu}
\affiliation{Center for Quantum Mathematics and Physics (QMAP),
	Department of Physics \& Astronomy,
	University of California, Davis, CA 95616, USA}

\date{\today}

\begin{abstract}
Using Hamiltonian methods in open bosonic string field theory, we derive the mass of a D0-brane as the central charge of the spontaneously broken Poincar{\'e} algebra in 26 dimensions. Though the formulas originate from string field theory, the calculation can be carried out completely on-shell. This illustrates a general procedure for computing spacetime central charges in string theory.
\end{abstract}

\maketitle

\pdfsecit{Introduction}---D-brane tension is perhaps the most elementary physical quantity computed in string field theory. As is well-known, it is the depth of the potential at the endpoint of open string tachyon condensation \cite{Sen,Taylor,Gaiotto,Kishimoto,Kudrna,Schnabl}. It can also be computed from the boundary state through the relation to Ellwood's gauge invariant observable~\cite{Ellwood,Kudrna2,Kudrna3}. In this Letter we present another computation of tension using the Hamiltonian formalism recently developed in \cite{Bernardes,Bernardes2,Bernardes3,Bernardes4,Bernardes5,Bernardes6,Ali}. The tension (or in this case mass) is determined as the central charge of the spontaneously broken Poincar{\'e} algebra on a D0-brane. The calculation is very simple and illustrates more broadly how target space central charges can be derived in string theory. 

{\it Spontaneously broken Poincar{\'e} algebra}---We consider Witten's open bosonic string field theory \cite{Witten} on a D0-brane in flat spacetime. The action~is
\begin{equation}S = -\frac{1}{g^2}\left[\frac{1}{2}\langle \Psi,Q\Psi\rangle + \frac{1}{3}\langle\Psi,\Psi*\Psi\rangle\right],\end{equation}
where $\Psi$ is a ghost number 1 state in the worldsheet theory of the D0-brane, $Q$ is the associated BRST operator, $\langle\cdot,\cdot\rangle$ is the BPZ inner product, $*$ is Witten's open string star product, and $g$ is the coupling constant. See \cite{Taylor2,Erler3} for review. Though the worldline of the D0-brane is 1-dimensional, the string field theory still possesses the full Poincar{\'e} invariance of the 26-dimensional background spacetime. The symmetry however is spontaneously broken. The Goldstone bosons for the broken symmetry are the transverse scalars on the D0-brane worldline. We consider a subalgebra of the Poincar{\'e} algebra consisting of translations and boosts along a spatial coordinate $x^1$ and translations in time $t$:
\begin{equation}
	p = \frac{\d}{\d x^1},\ \ \ \ j = x^1\frac{\d}{\d t} + t\frac{\d}{\d x^1},\ \ \ \ h = \frac{\d}{\d t}.
\end{equation}
The nonvanishing commutators are
\begin{equation}[p,j]= h,\ \ \ \ [h,j]=p.\label{eq:Poincare}\end{equation}
These generators are realized as transformations of the string field $\Psi$ which leave the string field theory action invariant. The transformations take the general form
\begin{align}
	\L_p\Psi & = p_0 + p_1\Psi + p_2(\Psi,\Psi)+p_3(\Psi,\Psi,\Psi)+\cdots,\nonumber\\
	\L_j\Psi & = j_0 + j_1\Psi + j_2(\Psi,\Psi)+j_3(\Psi,\Psi,\Psi)+\cdots,\nonumber\\
	\L_h \Psi & = h_1 \Psi, \label{eq:Psi_trans}
\end{align}
where $p_0,p_1,p_2,...$ and $j_0,j_1,j_2,...$ are an appropriate set of higher order products describing spatial translations and boosts and 
\begin{equation}h_1 = \oint \frac{d\xi}{2\pi i} \d X^0(\xi)\end{equation}
represents time translations. The string field theory action is manifestly time independent so time translation symmetry is simple to describe. The spatial translations and boosts are more complicated. They start with zero-products $p_0$ and $j_0$ which represent a shift in expectation value for a transverse scalar. This is how the change of the D0-brane's position is seen from the point of view of its own worldline. The shift by $p_0$ and $j_0$ alone is not a symmetry of the action because the BRST operator is transformed into the kinetic operator around a nontrivial string field. The higher products $p_1,p_2,...$ and $j_1,j_2,...$ must be constructed to map the kinetic operator back into the BRST operator of the D0-brane. This can be done following Sen and Zwiebach's proof of background independence \cite{Sen2,Sen3}, or using the intertwining solution of \cite{Erler,Erler2}. In any case the transformations should satisfy 
\begin{equation}
	[\L_p,\L_j] = -\L_h,\ \ \ \ \ [\L_h,\L_j] = -\L_p,
\end{equation}
so that the string field lives in a representation of the symmetry algebra. The sign appears because the commutator of field transformations implements the commutator of the differential operators in the opposite order.

The results of \cite{Bernardes5} give a construction of conserved charges associated to symmetries of Witten's string field theory. The results of \cite{Bernardes6} allow us to compute their Poisson bracket. That means that from~\eq{Psi_trans} one can derive a momentum $P$, a boost charge $J$, and a Hamiltonian $H$ whose Poisson brackets realize the Poincar{\'e} symmetry algebra
\begin{equation}
	[P,J] = M+H,\ \ \ \ [H,J] = P.
\end{equation}
By convention, conserved charges vanish at the vacuum $\Psi=0$. But the charges that appear in the Poisson bracket algebra should vanish at the vacuum of unbroken symmetry, which presently is 26-dimensional flat space without any D0-brane. Relative to that vacuum, the D0-brane itself has an energy given by its mass $M$. This is why the Hamiltonian appears in the Poisson bracket algebra with the additive constant $M$. The constant is the central charge of the Poisson bracket algebra, and is given by the formula \cite{Bernardes6},
\begin{equation}M = \frac{1}{g^2}\langle p_0,[Q,\sigma]j_0\rangle,\label{eq:M}\end{equation}
where $\sigma$ is the {\it sigmoid}. It is a ghost number zero operator subject to boundary conditions in time,
\begin{equation}\lim_{t\to-\infty} \sigma = 0,\ \ \ \ \lim_{t\to\infty}\sigma = 1.\label{eq:bc}\end{equation}
The transition between 0 and 1 gives a generalized notion of time slice. The expression for the central charge is related to the mass formula for a boosted solution derived in \cite{Bernardes3}. 

\pdfsecit{Computing the central charge}---To compute the mass we need the states $p_0$ and $j_0$. They take the form 
\begin{equation}p_0 = \mathcal{N} \alpha_{-1}^1c_1|0\rangle,\ \ \ \ j_0 = \mathcal{N}\alpha_{-1}^1x^0 c_1|0\rangle,\end{equation}
where $|0\rangle$ is the $SL(2,\mathbb{R})$ vacuum and $x^0$ is the position zero mode of the timelike free boson. The state $p_0$ represents a constant expectation value for the transverse scalar. This describes the infinitesimal translation of the D0-brane. The state $j_0$ represents an expectation value for the transverse scalar which is increasing linearly in time. This describes the infinitesimal boost of the D0-brane. The normalization $\mathcal{N}$ is a well-known number which relates the the expectation value of the transverse scalar to the physical displacement of the D0-brane. It was first determined in \cite{Taylor2,Sen4} by computing the mass of a stretched string connecting parallel D-branes. We give an independent derivation based on the relation between the Ellwood invariant and boundary state~\cite{Ellwood,Kudrna2}:
\begin{equation}
	\langle \mathcal{V}|(c_0-\overline{c}_0)\big(|B_\Psi\rangle - |B_0\rangle\big) 
	= -4\pi i \langle I|\mathcal{V}(i,\overline{i})|\Psi\rangle,\label{eq:Ellwood}
\end{equation}
where $\mathcal{V}$ is an on-shell closed string vertex operator, $|B_\Psi\rangle$ is the boundary state corresponding to the solution $\Psi$, $|B_0\rangle$ is the boundary state of the perturbative vacuum $\Psi=0$, and $I$ is the identity string field. We consider a solution that represents transverse displacement of the D0-brane a distance $\eps$ in the $x^1$ direction. To leading order in $\eps$ the solution $\Psi$ will be
\begin{equation}\Psi = \mathcal{N}\eps\, \alpha_{-1}^1 c_1|0\rangle.\end{equation}
We take the closed string vertex operator as
\begin{equation}
	\mathcal{V}(z,\overline{z}) = c\overline{c} e^{ik\cdot X(z,\overline{z})},\ \ \ \ \ k^2=4.
\end{equation}
After conformal transformation the Ellwood invariant can be mapped to a correlation function on the upper half plane with D0-brane boundary conditions 
\begin{equation}
	\langle I|\mathcal{V}(i,\overline{i})|\Psi\rangle = i\sqrt{2}\mathcal{N}\eps\langle c\overline{c}e^{ik\cdot X(u,\overline{u})} c\d X^1(0)\rangle_\UHP^{\mathrm{D0}},
\end{equation}
where $u$ is the location of the closed string vertex operator in the upper half plane. Because both the solution and the closed string vertex operator are dimension zero primaries, the nature of the conformal transformation to the upper half plane does not enter into the result. Splitting into matter and ghost parts and evaluating the correlation function gives
\begin{align}
	&\langle I|\mathcal{V}(i,\overline{i})|\Psi\rangle \nonumber
	\\ 
	& = i\sqrt{2}\mathcal{N}\eps\times \langle c(u)c(\overline{u})c(0)\rangle^\mathrm{ghost}_{\mathbb{C}}
	\nonumber \\
	&\quad \quad \times \langle e^{ik\cdot X(u,\overline{u})} \d X^1(0)\rangle_\UHP^{\mathrm{matter,D0}}\nonumber\\
	& = i\sqrt{2}\mathcal{N}\eps\times (u-\overline{u})u\overline{u}
	\nonumber \\
	&\quad \quad \times\left[\frac{i}{2}\left(\frac{k_1}{u}-\frac{k_1}{\overline{u}}\right)(2\mathrm{Im}(u))^{-k^2/2}\frac{Z_\mathrm{D0}}{\mathrm{vol}(X^0)}2\pi\delta(k_0)\right]\nonumber\\
	& = -\frac{\mathcal{N}\eps k_1}{\sqrt{2}}\frac{Z_\mathrm{D0}}{\mathrm{vol}(X^0)}2\pi\delta(k_0),
\end{align}
where we assume that the D0 brane is placed at the origin in all spatial coordinates and use the on-shell condition for the closed string tachyon. The factor $Z_\mathrm{D0}$ is the disk partition function for the worldsheet theory of the D0-brane, and the volume of time $\mathrm{vol}(X^0)$ is intended to cancel the $2\pi\delta(k_0)$ when $k_0=0$. Next we compute the overlap with the boundary state representing the translated D0-brane:
\begin{align}
	&\langle \mathcal{V}|(c_0-\overline{c}_0)|B_\Psi\rangle 
	\nonumber \\
	& = \Big\langle\big[\d c c\overline{c}+c\d\overline{c}\overline{c} \big]e^{ik\cdot X(u,\overline{u})}\Big\rangle_\UHP^{\mathrm{translated\, D0}}\nonumber\\
	& = \big\langle \d c c(u) c(\overline{u}) +c(u)\d cc(\overline{u})\big\rangle_\mathbb{C}^\mathrm{ghost}
	\nonumber \\
	&\quad \quad \times\big\langle e^{ik\cdot X(u,\overline{u})}\Big\rangle_\UHP^{\mathrm{matter,\,translated\, D0}}\nonumber\\
	& = 2(u-\overline{u})^2\times \left[(2\mathrm{Im}(u))^{-k^2/2} e^{ik_1 \eps}\frac{Z_\mathrm{D0}}{\mathrm{vol}(X^0)}2\pi\delta(k_0)\right]\nonumber\\
	& = -2 e^{ik_1 \eps}\frac{Z_\mathrm{D0}}{\mathrm{vol}(X^0)}2\pi\delta(k_0).
\end{align}
The plane wave factor comes from the translated D0 brane boundary condition. Subtracting the result from $|B_0\rangle$ and expanding to leading order in $\eps$ gives
\begin{equation}
	\langle \mathcal{V}|c_0^-\big(|B_\Psi\rangle - |B_0\rangle\big) = -2i \eps k_1 \frac{Z_\mathrm{D0}}{\mathrm{vol}(X^0)}2\pi\delta(k_0)+ \mathcal{O}(\eps^2).
\end{equation}
Comparing to the Ellwood invariant through \eq{Ellwood} we find
\begin{equation}\mathcal{N} = -\frac{1}{\pi\sqrt{2}},\end{equation}
which agrees with the known result, see equation (234) in \cite{Taylor}.

We can now compute the central charge
\begin{align}
	M & = \frac{1}{g^2}\langle p_0,[Q,\sigma]j_0\rangle\nonumber\\
	& = \frac{1}{2\pi^2g^2} \langle 0|c_{-1}\alpha_1^1[Q,\sigma] x^0 \alpha_{-1}^1 c_1|0\rangle.\label{eq:cc}
\end{align}
We can already recognize the well-known tension formula in string field theory in the prefactor. The value of the central charge should be independent of the choice of sigmoid provided that the boundary conditions~\eq{bc} are satisfied. We choose a sigmoid made out of the zero mode of the timelike free boson
\begin{equation}\sigma = \int \frac{dE}{2\pi}\sigma(E) e^{i E x^0},\end{equation}
where the Fourier modes $\sigma(E)$ have a pole that goes as $1/iE$ at $E=0$. This is the momentum space expression of the boundary conditions \eq{bc}. The time derivative of the sigmoid is 
\begin{equation}\dot{\sigma} = \int \frac{dE}{2\pi}\dot{\sigma}(E) e^{i E x^0},\label{eq:21}\end{equation}
where the Fourier modes $\dot{\sigma}(E)$ satisfy
\begin{equation}\dot{\sigma}(0) = 1.\end{equation}
With this choice of sigmoid we can compute \cite{Bernardes2}
\begin{equation}
	[Q,\sigma] = \gamma^0\dot{\sigma} + \dot{\sigma}\gamma^0,\ \ \ \ \gamma^\mu = -\frac{i}{\sqrt{2}}\sum_{n\in\mathbb{Z}}c_n\alpha^\mu_{-n}.
\end{equation}
The oscillator part of $\gamma^0$ does not contribute to the central charge so the relevant formula is
\begin{equation}
	[Q,\sigma] = (ic_0 p_0)\dot{\sigma} + \dot{\sigma}(i c_0 p_0)+\mathrm{oscillators},
\end{equation}
where in this equation $p_0$ is the momentum zero mode of the timelike free boson. Plugging in to \eq{cc} the first term does not contribute because the momentum zero mode annihilates the vacuum, and in the second term the momentum zero mode deletes the factor of $x^0$. This leaves
\begin{equation}
	M = \frac{1}{2\pi^2 g^2} \langle 0|c_{-1}c_0 c_1\dot{\sigma}|0\rangle.
\end{equation}
Expanding $\dot{\sigma}$ as in \eq{21} the plane wave zero mode operator shifts the momentum of the $SL(2,\mathbb{R})$ vacuum
\begin{equation}|E\rangle = e^{iE x^0}|0\rangle,\end{equation}
so that
\begin{equation}
	M = \frac{1}{2\pi^2 g^2}\int \frac{dE}{2\pi}\dot{\sigma}(E) \langle 0|c_{-1}c_0 c_1|E\rangle.
\end{equation}
Allowing for a nontrivial normalization of the disk partition function we have
\begin{equation}
	\langle 0|c_{-1}c_0 c_1|E\rangle = \frac{Z_\mathrm{D0}}{\mathrm{vol}(X^0)}2\pi\delta(E),
\end{equation}
and
\begin{align}
	M & = \frac{1}{2\pi^2 g^2}\int \frac{dE}{2\pi}\dot{\sigma}(E) \frac{Z_\mathrm{D0}}{\mathrm{vol}(X^0)}2\pi\delta(E)\nonumber\\
	& = \frac{1}{2\pi^2 g^2} \frac{Z_\mathrm{D0}}{\mathrm{vol}(X^0)},
\end{align}
where we used $\dot{\sigma}(0)=1$. This is the correct value for the D-brane tension \cite{Sen4}.

\pdfsecit{Discussion}---While the motivation comes from string field theory, it is interesting that all calculations could be carried out on-shell. We never needed to define string vertices and calculate conformal transformations of off-shell vertex operators. It should be possible to extend this technique to determine Ramond-Ramond charges from the spontaneously broken supersymmetry algebra in superstring field theory. More ambitiously, we could hope to derive  formulas for Ramond-Ramond charges as topological invariants of the string field algebra, perhaps using wedge-based analytic methods \cite{Erler,Erler4,Erler5} in the Wess-Zumino-Witten-like superstring field theory \cite{Berkovits,Kunitomo}. Another interesting application is to compute the Brown-Henneaux central charge~\cite{Brown} for the spacetime Virasoro symmetry of strings in $AdS_3$ \cite{Giveon,Boer,Ooguri}.

\pdfsecit{Acknowledgments}---We thank Ashoke Sen for comments on the draft. AHF thanks Mukund Rangamani for conversations. The work of VB and TE was supported by the European Structural and Investment Funds and the Czech Ministry of Education, Youth and Sports (project No. FORTE—CZ.02.01.01/00/22\_008/0004632). The work of AHF is supported by the U.S. Department of Energy, Office of Science, Office of High Energy Physics of U.S. Department of Energy under grant Contract Number DE-SC0009999, and the funds from the University of California.

\end{document}